\documentstyle[12pt]{article}

\advance \textheight by \topskip
\makeatletter
\def\eqnarray{\stepcounter{equation}\let\@currentlabel=\theequation
\global\@eqnswtrue
\global\@eqcnt\z@\tabskip\@centering\let\\=\@eqncr
$$\halign to \displaywidth\bgroup\@eqnsel\hskip\@centering
  $\displaystyle\tabskip\z@{##}$&\global\@eqcnt\@ne 
  \hfil$\displaystyle{\hbox{}##\hbox{}}$\hfil
  &\global\@eqcnt\tw@ $\displaystyle\tabskip\z@
  {##}$\hfil\tabskip\@centering&\llap{##}\tabskip\z@\cr}
\@addtoreset{equation}{section}
  \def\theequation{\thesection.\arabic{equation}}
\makeatother

\mathchardef\by="0202

\begin{document}

\title{\bf Three aspects of bosonized supersymmetry
and linear differential field equation with reflection}

\author{Jorge Gamboa${}^{a}$\thanks{E-mail:
jgamboa@lauca.usach.cl}, 
Mikhail Plyushchay${}^{a,b}$\thanks{E-mail: mplyushc@lauca.usach.cl}, 
Jorge Zanelli${}^{a,c}$\thanks{E-Mail: jz@cecs.cl}\\
\\
{\small ${}^{a}${\it Departamento de F\'{\i}sica, 
Universidad de Santiago de Chile,}}
{\small {\it Casilla 307, Santiago 2, Chile}}\\
{\small ${}^{b}${\it Institute for High Energy Physics, Protvino,
Russia}}\\
{\small ${}^{c}${\it Centro de Estudios Cient\'{\i}ficos
de Santiago,}}
{\small {\it Casilla 16443, Santiago, Chile}}}
\date{}

\maketitle
\vskip-1.0cm

\begin{abstract}
Recently it was observed  by one of the authors that supersymmetric
quantum mechanics (SUSYQM) admits a formulation in terms of only one
bosonic degree of freedom. Such a construction, called the minimally
bosonized SUSYQM, appeared in the context of integrable systems and
dynamical symmetries. We show that the minimally bosonized SUSYQM can be
obtained from Witten's SUSYQM by applying to it a nonlocal unitary
transformation with a subsequent reduction to one of the eigenspaces of
the total reflection operator. The transformation depends on the parity
operator, and the deformed Heisenberg algebra with reflection, intimately
related to parabosons and parafermions, emerges here in a natural way. It
is shown that the minimally bosonized SUSYQM can also be understood as
supersymmetric two-fermion system.  With this interpretation, the
bosonization construction is generalized to the case of $N=1$
supersymmetry in 2 dimensions.  The same special unitary transformation
diagonalises the Hamiltonian operator of the $2D$ massive free Dirac
theory. The resulting Hamiltonian is not a square root like in the
Foldy-Wouthuysen case, but is linear in spatial derivative. Subsequent
reduction to `up' or `down' field component gives rise to a linear
differential equation with reflection whose `square' is the massive
Klein-Gordon equation. In the massless limit this becomes the self-dual
Weyl equation. The linear differential equation with reflection admits
generalizations to higher dimensions and can be consistently coupled to
gauge fields.  The bosonized SUSYQM can also be generated applying the
nonlocal unitary transformation to the Dirac field in the background of a
nonlinear scalar field in a kink configuration.  

\vskip1mm
\noindent 
{\it PACS:} 11.30.Pb, 11.10.Lm, 03.65.Pm, 11.10.Kk

\noindent
{\it Keywords:} Supersymmetric quantum mechanics; Bosonization;
Nonlocality; Fermionic zero modes;
Linear differential field equation with reflection
\end{abstract}

\newpage

\section{Introduction}

Supersymmetric quantum mechanics (SUSYQM) was introduced by
Witten \cite{wit} as a toy model for studying supersymmetry
breaking mechanism that would solve the hierarchy problem.
Since then, SUSYQM has found many important applications.  Their
exhaustive list alongside with corresponding references can be
found, e.g., in the review papers \cite{d2,susyqm}.  A new
aspect of SUSYQM,  recently discussed in the context of
integrable systems and dynamical symmetries, is its minimally
bosonized form \cite{mp}.  The name of the construction stems
from its formulation in terms of only one bosonic degree of
freedom.  The supercharge and Hamiltonian operators of the
minimally bosonized supersymmetry are given by
\[
Q_1=-\frac{i}{\sqrt{2}}\left(\frac{d}{dx}+W_-(x)R\right),\quad
Q_2=iRQ_1,\quad 
H=\frac{1}{2}\left(-\frac{d^2}{dx^2}+W_-^2-W'_- R\right), 
\]
where $W_-(x)=-W_-(-x)$ is the odd superpotential, and $R$ is
the reflection operator, $R^2=1$, $Rx=-xR$. It was shown in
\cite{mp} that the construction can in principle be extended to
the case of $OSp(2\vert 2)$ supersymmetry and that it is related
to the Witten's SUSYQM in a nontrivial way, but the exact form
of the relationship was not established.

In this paper we investigate the relationship between
conventional and bosonized forms of SUSYQM and discuss two other
aspects of bosonized supersymmetry: its interpretation as
supersymmetry of a system of identical fermions and as a
symmetry in a system of a Dirac field in the background of a
nonlinear scalar field.  The bosonization construction applied
to the $2D$ massive Dirac field allows us also to obtain a
linear differential field equation with reflection whose 
`square' is the Klein-Gordon equation.

The paper is organized as follows.  In Section 2 we find the
nonlocal unitary transformation and subsequent reduction
procedure relating the conventional form of Witten's SUSYQM with
the bosonized form.  Here we discuss general properties of the
bosonized supersymmetry and arrive naturally at the deformed
Heisenberg algebra with reflection emerging in some integrable
systems \cite{poly,macf} and closely related to parabosons and
parafermions \cite{para,def}.  In Section 3 we show that
bosonized supersymmetry can be understood as a supersymmetry of
the system of two identical fermions.  This interpretation
allows us to generalize the bosonization construction to the
case of $N=1$ supersymmetry in two spatial dimensions.  In
Section 4 we find that the same nonlocal unitary transformation,
which relates the conventional supersymmetry with the bosonized
one, diagonalises the Hamiltonian of the $(1+1)$-dimensional
massive Dirac equation.  Subsequent reduction of the transformed
Dirac equation supplies us with the linear differential equation
with reflection whose `squared form' is the Klein-Gordon
equation.
In the massless limit this becomes the self-dual Weyl equation.
We generalize the obtained linear equation to
higher (arbitrary) dimensions and show that it admits switching
on gauge interactions of special form.  Finally, we observe how
the bosonized SUSYQM appears under application of the same
nonlocal unitary transformation to the $2D$ system of Dirac
field in the background of a soliton.  Here the corresponding
phases of the bosonized supersymmetry signal the presence or
absence of fermionic zero modes.  We conclude in Section 5
giving the list of open problems for further investigation.

\section{Special unitary transformation and bosonized SUSY}

In this Section we show how the minimally bosonized supersymmetric
quantum
mechanics \cite{mp} can be related to 
Witten's SUSYQM \cite{wit}
by applying to the latter the special nonlocal unitary
transformation with a subsequent reduction
to one of the eigenspaces of the total reflection operator.
The form of such bosonization procedure is found by analyzing
the structure of the simplest supersymmetric quantum mechanical system
(superoscillator) and then is employed for the 
general case of $N=1$ supersymmetry.

\subsection{Simplest bosonized supersymmetric system}

Let us consider the superoscillator \cite{nic}
given by the Hamiltonian 
\begin{equation}
H=\frac{1}{2}\{b^{+},b^{-}\}+
\frac{1}{2}[f^{+},f^{-}],
\label{h1}
\end{equation}
where $b^\pm$ and $f^\pm$
are the bosonic and fermionic creation-annihilation operators,
\[
[b^{-},b^{+}]=1,
\quad
\{f^{-},f^{+}\}=1,\quad
f^{+2}=f^{-2}=0,
\quad
[b^{\pm},f^{\pm}]=0.
\]
The supercharge operators
\begin{equation}
Q_{\pm}=b^\mp f^\pm
\label{schar}
\end{equation}
are dynamical integrals of motion forming together with
the Hamiltonian (\ref{h1})
the superalgebra of $N=1$ supersymmetry:
\begin{equation}
\{Q_+,Q_-\}=H,\quad
Q_+^{2}=Q_-^{2}=0,\quad
[H,Q_\pm]=0.
\label{s2}
\end{equation}
The Hamiltonian
is the sum of the bosonic and fermionic number operators,
$H=N_b+N_f$, $N_b=b^+b^-$, 
$[N_b,b^\pm]=\pm b^\pm$, $N_f=f^+f^-$, 
$[N_f,f^\pm]=\pm f^\pm$. Thus,
in the basis $\vert n_b,n_f\rangle$,
$N_b\vert n_b,n_f\rangle=n_b\vert n_b,n_f\rangle$,
$N_f\vert n_b,n_f\rangle=n_f\vert n_b,n_f\rangle$,
$n_b=0,1,\ldots$, $n_f=0,1$,
$H$ is diagonal and its spectrum is
\begin{equation}
E_n=\left[\frac{n+1}{2}\right],
\label{spe0}
\end{equation}
where $n=2n_b+n_f=0,1,\ldots$, and $[.]$ means the integer part.
The state $\vert 0,0\rangle$ is the 
SUSY singlet corresponding to $E_0=0$,
whereas all other states are SUSY doublets
with $E_{2k-1}=E_{2k}>0$, $k=1,\ldots$.
The operator
\[
F=[f^+,f^-]=-(-1)^{N_f},
\]
satisfies the relations
$
\{F,f^\pm\}=0$, 
$F^2=1.$ $F$ is the grading operator, which
may also be interpreted as fermionic reflection operator.
Since $N_f=\frac{1}{2}(1+F)$, instead of $n_f$,
one can characterize the complete basis by 
the eigenvalues
of the operator $F$: $F\vert n_b,\varphi\rangle=
\varphi\vert n_b,\varphi\rangle$, $\varphi=\pm 1$.
The operators  $f^\pm$ and $F$ may be represented with 
Pauli matrices,  
$
f^\pm=\frac{1}{2}(\sigma_1\pm i\sigma_2),
$
$
F=\sigma_3.
$

Let us  rearrange the entire spectrum:
\begin{equation}
E_n=E^+_n\oplus E^-_n,\quad
E^+_n=2\left[\frac{n+1}{2}\right],\quad
E^-_n=2\left[\frac{n}{2}\right]+1,\quad
n=0,1,\ldots.
\label{spe1}
\end{equation}
For $n\neq0$, $E^+_n$ are doubly degenerate, but
$E^+_0=0$ is nondegenerate.
On the  other hand, the spectrum $E^-_n$ is doubly degenerate
for all $n$.
Thus, $E^+_n$ is the spectrum
of a system with unbroken SUSY, whereas
$E^-_n$ corresponds to some another system in a phase of spontaneously
broken SUSY. The corresponding eigenstates are
\[
\vert n,\epsilon\rangle\equiv
(\vert n_b=2n,\varphi=-\epsilon\cdot 1\rangle,
\vert n_b=2n+1,\varphi=\epsilon\cdot 1\rangle),\quad
H\vert n,\epsilon\rangle=E^\epsilon_n
\vert n,\epsilon\rangle.
\]
The subspaces with $\epsilon=+$
and $\epsilon=-$ are separated by the operator
${\cal R}=-RF$, ${\cal R}\vert n,\epsilon\rangle
=\epsilon\vert n,\epsilon\rangle$,
where $R=(-1)^{N_b}$ is the reflection (parity)
bosonic operator, $\{R,b^\pm\}=0$, $R^2=1$.
${\cal R}$ represents the total reflection
operator, $\{{\cal R},b^\pm\}=0$,
$\{{\cal R},f^\pm\}=0,$
${\cal R}^2=1$, and there is a
unitary transformation relating  ${\cal R}$ to
the fermionic reflection operator $F=\sigma_3$,
$U\sigma_3U^\dagger={\cal R}$. The
corresponding unitary operator may be written in the form
\begin{equation}
U=\exp(i\pi{\cal S}_-\Pi_+)=
{\cal S}_+-R{\cal S}_-,
\label{unop}
\end{equation}
where ${\cal S}_\pm=
\frac{1}{2}(1\pm\sigma_1)$
and 
$\Pi_\pm=\frac{1}{2}(1\pm R)$
are the projector operators,
${\cal S}_\pm^2={\cal S}_\pm$,
${\cal S}_+{\cal S}_-=0$,
${\cal S}_++{\cal S}_-=1$,
$\Pi_\pm^2=\Pi_\pm$, $\Pi_+\Pi_-=0$, $\Pi_++\Pi_-=1$.
{}From (\ref{unop}) one can see that 
$U^\dagger=U^{-1}=U$, and calling 
$A'=UAU^{-1}$, then
\begin{equation}
\sigma_1'=\sigma_1,\quad
\sigma_2'=-R\sigma_2,\quad
\sigma'_3=-R\sigma_3={\cal R},\quad
b^\pm{}'=b^\pm\sigma_1\equiv a^\pm.
\label{top}
\end{equation}
$U$ transforms eigenstates of $F$ into eigenstates
of ${\cal R}$: 
$\vert n,\epsilon\rangle=U\vert n,\varphi\rangle$.
The initial ({\it untransformed}) supercharges and Hamiltonian
can be written in terms of the {\it transformed} 
operators $a^\pm$, ${\cal R}$ as 
\[
Q_\pm=a^\mp\frac{1}{2}(1\pm R{\cal R}),\quad
H=\frac{1}{2}\{a^+,a^-\}-\frac{1}{2}R{\cal R},
\quad
R=(-1)^{N_a}, 
\]
where $N_a\equiv a^+a^-$, and
we have taken into account that $N_b=N_a$. 

Let us consider now the restriction of these
supersymmetry generators 
to the eigenspaces of 
the operator ${\cal R}$.
We have $Q_\pm\vert n,\epsilon\rangle=Q_\pm^\epsilon
\vert n,\epsilon\rangle$,
$H\vert n,\epsilon\rangle=H^\epsilon\vert n,\epsilon\rangle$,
where 
\begin{equation}
Q_\pm^\epsilon=a^\mp\Pi_{\pm\epsilon},\quad
H^\epsilon=\frac{1}{2}\{a^+,a^-\}-\frac{1}{2}\epsilon R.
\label{bosqh}
\end{equation}
Since the operators $Q_\pm$ and $H$ commute with
${\cal R}$,
their restrictions (\ref{bosqh}) satisfy
the same $N=1$ superalgebra (\ref{s2}).
Thus, reducing the system to the subspaces
with $\epsilon=+$ or $\epsilon=-$
does not affect the supersymmetry algebra.
As a result, we find two systems described only
by the bosonic operators $a^\pm$ and given by the 
Hamiltonians $H^\epsilon$, $\epsilon=+,-$. 
In the case of $\epsilon=+$ the 
system exhibits the spectrum $E^+_n$
corresponding to the exact supersymmetry,
whereas the choice $\epsilon=-$ gives 
the system in the phase
of spontaneously broken supersymmetry
with the spectrum $E^-_n$.
The operators $Q_\pm^\epsilon$ given in terms of only
bosonic operators $a^\pm$ are the corresponding supercharge
operators.
Thus, bosonized supersymmetry
in the exact or spontaneously broken phases can be produced
via the unitary transformation  
(\ref{unop})
with a subsequent reduction of the system
to one of the eigenspaces of the total reflection operator.

The same result may be obtained applying the 
unitary transformation $U$ 
to the supercharges and Hamiltonian 
(since $U=U^{-1}$) with the subsequent reduction
of the system to the corresponding eigenspaces of the
operator $F$.
Indeed, the transformed operators $Q'_\pm$ and
$H'$ take the form
\begin{equation}
Q'_{\pm}=b^\mp\frac{1}{2}(1\pm R\sigma_3),\quad
H'=\frac{1}{2}\{b^+,b^-\}-\frac{1}{2}R\sigma_3,
\label{qhtr}
\end{equation}
and commute with the {\it untransformed}
operator $F=\sigma_3$. Restricting 
these operators on the eigenspaces of $F$
with $\varphi=+1$ or $-1$, and then changing the notation 
$\varphi\rightarrow \epsilon$,
$b^\pm\rightarrow a^\pm$,
we arrive at the bosonized supercharge and Hamiltonian
operators (\ref{bosqh}). This alternative way is more convenient
for generalizations.

In the coordinate representation, where
$b^\pm=\frac{1}{\sqrt{2}}(x\mp \frac{d}{dx})$,
the  operator $R$ acts as 
$R\psi(x)=\psi(-x)$,
and we see that the operator (\ref{unop})
generates the unitary transformation
\[
\Psi(x)\rightarrow \Psi'(x)=U\Psi(x)=
{\cal S}_+\Psi(x)-{\cal S}_-\Psi(-x),
\]
which is {\it nonlocal}:
the transformed state at point $x$ is a linear combination 
of initial states taken at points $x$ and $-x$.

\subsection{Bosonized SUSYQM: general case}

We now generalize the bosonization construction
starting with an 
arbitrary supersymmetric
quantum mechanical system. 
Consider the SUSYQM system characterized by 
the supercharges $Q_\pm$ and the Hamiltonian 
$H$ defined by the superpotential $W(x)$ \cite{wit}:
\begin{equation}
Q_\pm=\frac{1}{\sqrt{2}}\left(\pm \frac{d}{dx}
+W(x)\right)\cdot \frac{1}{2}(\sigma_1\pm i\sigma_2),
\quad
H=\frac{1}{2}\left(-\frac{d^2}{dx^2}+W^2(x)+W'(x)\sigma_3\right).
\label{wh}
\end{equation}
Let us decompose $W(x)$ into even and
odd parts, $W(x)=W_+(x)+W_-(x)$,
$W_\pm(-x)=\pm W_\pm(x)$, and
realize the unitary transformation with the operator
(\ref{unop}).
Under it, the operators (\ref{wh}) are transformed into
\[
Q'_\pm=\frac{1}{\sqrt{2}}\left(
\pm\frac{d}{dx}+W_+(x)\sigma_1+W_-(x)\right)\frac{1}{2}
(1\pm\sigma_3 R),
\]
\[
H'=\frac{1}{2}\left(
-\frac{d^2}{dx^2}+(W_+(x)\sigma_1+W_-(x))^2-
(W_+'(x)\sigma_1 +W_-'(x))\sigma_3 R\right).
\]
If the even part
of the superpotential vanishes, $W_+=0$,
(as in the case of the superoscillator considered above,
$W(x)=W_-(x)=x$),
the operators $Q'_\pm$ and $H'$
commute with $\sigma_3$
and do not mix `up' and `down' states.
Restricting these operators on the
`up' ($\epsilon=+$) and `down' 
($\epsilon=-$) eigenspaces of $\sigma_3$,
we get
\[
Q_\pm^\epsilon= \frac{1}{\sqrt{2}}\left(\pm
\frac{d}{dx}+W_-(x)\right)\cdot \Pi_{\pm\epsilon}, 
\quad
H^\epsilon=\frac{1}{2}
\left(-\frac{d^2}{dx^2}+W_-^2(x)-\epsilon W'_-(x)R\right),
\]
that generalizes the simplest bosonized supersymmetric
system (\ref{bosqh}).
The restricted operators 
$Q_\pm^\epsilon$ and $H^\epsilon$
form the $N=1$ superalgebra (\ref{s2}) both for
$\epsilon=+$ and $\epsilon=-$.
These two 
cases are related in a simple way:
$
Q_\pm^\epsilon(-W_-)=-Q_\mp^{-\epsilon}(W_-),
$
$
H^\epsilon(-W_-)=H^{-\epsilon}(W_-).
$
Therefore, the general case of the bosonized SUSYQM can be given by
the hermitian supercharge and Hamiltonian operators
\begin{equation}
Q_1=
-\frac{i}{\sqrt{2}}\left(\frac{d}{dx}+W_-R\right),
\quad
Q_2=iRQ_1,
\quad
H=
\frac{1}{2}\left(-\frac{d}{dx^2}+W_-^2-W_-'R\right),
\label{hq12}
\end{equation}
containing arbitrary odd function (superpotential) $W_-(x)$ and
satisfying the superalgebra $\{Q_i,Q_j\}=2\delta_{ij}H$,
$[H,Q_i]=0$, $i,j=1,2$.
The hermitian supercharges are related to $Q_\pm^+$
as $Q_1=i(Q_-^+-Q_+^+)$, $Q_2=Q_+^++Q_-^+$.

The generators of bosonized supersymmetry 
act on the space of wave functions
$\Psi(x)$,
where the action of 
reflection operator $R$ is defined by $R\Psi(x)=\Psi(-x)$.
It should be noted that
the operator $i\sqrt{2}Q_1$ is related to 
$\frac{d}{dx}$ through
\[
i\sqrt{2}Q_1\exp(-\omega_+R)=
\exp(+\omega_+R)\frac{d}{dx}
\]
involving the nonunitary operator $\exp(\omega_+R)$,
$\omega_+(x)=\int^xW_-(y)dy$.
The argument of the exponent is operator-valued,
and it can be verified that
\[
-2Q_1^2\exp(-\omega_+R)\Psi(x)=\exp(-\omega_+R)(\frac{d}{dx}+2W_-(x))
\frac{d}{dx}\Psi(x).
\]
This means that the action of $i\sqrt{2}Q_1$  
is not reducible to the simple derivative.

Let us also note that for the particular case,
$W_-(x)=-\frac{\nu}{2x}$,
the operator $i\sqrt{2}Q_1$  
coincides with the
Yang-Dunkl operator \cite{yd}
$
D_\nu=\frac{d}{dx}-\frac{\nu}{2x}R
$
which occurs in the Calogero model \cite{poly,macf},
where $R$ is an exchange operator.
With the extended differential operator $D_\nu$,
we arrive at the deformed Heisenberg algebra
with reflection 
\begin{equation}
[a^-,a^+]=1+\nu R,\quad
R^2=1,\quad
\{a^\pm,R\}=0,
\label{aad}
\end{equation}
with $a^\pm=\frac{1}{\sqrt{2}}(x\mp iD_\nu)$.
This algebra is intimately related
to parabosons \cite{para} and parafermions \cite{def},
and to the $osp(1\vert2)$ and
$osp(2\vert 2)$ superalgebras \cite{mp,def}.

Let us now decompose the wave function 
into even and odd parts, $\Psi(x)=\Psi_+(x)+\Psi_-(x)$,
$\Psi_\pm(-x)=\pm\Psi(x)$,
and define the differential operators
\begin{equation}
A^\pm=\frac{1}{\sqrt{2}}\left(\mp \frac{d}{dx}+W_-\right).
\label{AA}
\end{equation}
We have 
$[A^-,A^+]=W_-'$, 
$
Q_1=i(A^+\Pi_- - A^-\Pi_+),
$
and
$
H=A^+A^-\Pi_+ + A^-A^+\Pi_-
$
where $\Pi_+$ and $\Pi_-$ are projectors on the 
even and odd subspaces, $\Pi_\pm\Psi(x)=\Psi_\pm(x)$.
Then the eigenvalue problems 
$Q_1\Psi(x)=\lambda\Psi(x)$, $H\Psi(x)=\lambda^2\Psi(x)$
are decomposed into
\begin{equation}
iA^-\Psi_+=\sqrt{2}\lambda\Psi_-,
\quad
-iA^+\Psi_-=\sqrt{2}\lambda\Psi_+,
\label{c1}
\end{equation}
and 
\begin{equation}
A^+A^-\Psi_+=2\lambda^2\Psi_+,\quad
A^-A^+\Psi_-=
2\lambda^2\Psi_-,
\label{c2}
\end{equation}
respectively.
In the conventional form of SUSYQM,
the eigenvalue problems for the supercharge
$Q_1=\frac{i}{\sqrt{2}}\sigma_1(\frac{d}{dx}+W\sigma_3)$
and Hamiltonian
$H=\frac{1}{2}(-\frac{d^2}{dx^2}+W^2-W'\sigma_3)$
are formally given
by the same relations (\ref{c1})
and (\ref{c2}), but 
$\Psi_\pm$ in that case
being the `up' and `down'
components, with no parity restrictions
imposed on them or on the superpotential $W$. 
Therefore, the bosonization procedure is reduced {\it formally} to the
change  of the
superpotential $W(x)$ (having no parity restrictions
in the general case of Witten's SUSYQM) for odd
superpotential
and to the change of `up' and `down' quantum states of the
conventional SUSYQM for even and odd wave functions in the bosonized
version of SUSYQM.
As we shall see below,
these parity restrictions
on the superpotential and wave functions
imply that the bosonized SUSYQM
can be understood as a supersymmetric
system of two identical fermions.

In correspondence with Eqs. (\ref{c1}), (\ref{c2}),
if the bosonized supersymmetric system 
is in the phase of spontaneously unbroken
supersymmetry,
its vacuum state has to satisfy 
either  $A^-\Psi_+=0$ or $A^+\Psi_-=0$.
Solutions to these equations are
$\Psi_+^{(0)}=N_+\exp(-\int^x W_-(y)dy)$,
$\Psi_-^{(0)}=N_-\exp(+\int^x W_-(y)dy)$.
The second solution would be odd
only for $N_-=0$.
Therefore, if the superpotential $W_-$ is
such that $\Psi_+^{(0)}$ is normalizable,
it is the case of exact bosonized SUSY,
i.e. the supersymmetric vacuum state
(if it exists) is always described by an even function.
In the bosonized version,
like the case of conventional SUSYQM,
the SUSY partner eigenfunctions belonging to the same
energy eigenvalue $E>0$ are related by the operators (\ref{AA}),
$\Psi_-\propto A^-\Psi_+$,
$\Psi_+\propto A^+\Psi_-$.

Examples of the systems with unbroken bosonized 
supersymmetry are provided by
the following superpotentials:
\[
W_-=\epsilon x^{2k+1},\, \, k=0,1,\ldots,\quad
W_-=\epsilon \frac{x}{\vert x\vert},\quad
W_-=\epsilon \sinh x,\quad
W_-=\epsilon \tanh x,\quad
\epsilon=+.
\]
A superpotential of the form 
$W_-=\epsilon x-\frac{\nu}{2x}$
gives rise to exact SUSY when $\epsilon=+$ and $\nu>-1$.
The supercharge and Hamiltonian
operators for this 
system can be represented in the form (\ref{bosqh})
in terms of creation-annihilation
operators of the deformed Heisenberg algebra with reflection
(\ref{aad}).
This system turns out to be isospectral to the
bosonized superoscillator in the phase of exact supersymmetry
(with $E^+_n=2[(n+1)/2]$) discussed in the previous subsection.

On the other hand, for $\epsilon=-$
the system with superpotential $W_-=\epsilon x-\frac{\nu}{2x}$
is in the phase of the spontaneously broken
supersymmetry with the spectrum 
$E^-_n=2[n/2]+1+\nu$.
In this case the deformation
parameter defines the scale of supersymmetry breaking
(for details, see ref. \cite{mp}).

\section{SUSY of two-fermion system and bosonized SUSY}
As we saw,
the bosonized form of SUSYQM
can be formally  obtained by 
imposing the corresponding parity restrictions
on the superpotential and `up' and `down' 
states of the conventional SUSYQM.
But analogous parity restrictions 
on the physical operators 
and wave functions  
emerge when a system of identical fermions
is described.
Having in mind this observation,
here we show that the minimally bosonized supersymmetric quantum mechanics
can be understood as a supersymmetric two-fermion system.
With this interpretation,
the bosonization construction is generalized 
to the case of $N=1$ supersymmetry 
in two spatial dimensions.

Let us consider a
system of two {\it identical} fermions on the line.
It can be described by the wave function
$\Psi_{s_1,s_2}(x_1,x_2)=
-\Psi_{s_2, s_1}({x}_2,{x}_1)$, 
where indices $s_1$, $s_2$ correspond to spin degrees
of freedom of the particles, and
we assume that unlike the
$1$-dimensional coordinate space, 
the spin operator space is $3$-dimensional\footnote{Let us indicate 
the difference of the present case
of fermions with spin from the 
systems of spinless fermions on the line,
where the problem of self-adjointness 
of physical operators is essential \cite{f1,f2,gz}.}.
It is convenient to pass over to 
the center of mass and relative coordinates,
$X=\frac{1}{2}({x}_1+{x}_2)$,
${x}={x}_1-{x_2}$,
as well as to the basis of the total spin,
$J_i=\frac{1}{2}(\sigma_i\otimes 1+1\otimes\sigma_i)$,
$i=1,2,3$. Then, omitting 
the dependence on center of mass coordinate,
we describe the two-fermion system 
by the wave functions of the form
\begin{equation}
\Psi_f({x})=\chi^{j_3}_s\psi^{j_3}_-({x})
+\chi_a\psi_+({x}), 
\label{gen}
\end{equation}
where
$j_3=+1,0,-1$.
Here
$\chi^{+1}_s=\vert +\rangle\vert+\rangle$,
$\chi^{-1}_s=\vert -\rangle\vert-\rangle$ and
$\chi^{0}_s=\frac{1}{\sqrt 2}(\vert +\rangle\vert-\rangle
+\vert -\rangle\vert+\rangle)$
are symmetric spin states forming a vector triplet,
$J_iJ_i\chi_s^{j_3}=2\chi_s^{j_3}$, 
$J_3\chi^{j_3}_s=j_3\chi^{j_3}_s$, and
$\chi_a=\frac{1}{\sqrt{2}}(\vert +\rangle\vert-\rangle
-\vert -\rangle\vert+\rangle)$ is antisymmetric
spin-0 singlet state, $J_iJ_i\chi_a=J_3\chi_a=0$;  
$\psi^{j_3}_-$ are odd functions,
$\psi^{j_3}_-(-{x})=
-\psi^{j_3}_-({x})$, whereas $\psi_+$ is an even
function,
$\psi_+(-{x})=\psi_+({x})$.
To simplify the notation, below we denote 
$\chi^{\pm 1}_s$, 
$\chi^0_s$ and $\psi^0_-$
by $\chi^\pm$, 
$\chi_s$ and $\psi_-$, respectively.

We want to realize $N=1$ supersymmetry
on this system of two identical fermions.
This means that 
our task is to write
the supercharge and Hamiltonian operators  
in a form 
similar to that in Witten's SUSYQM: 
\begin{equation}
Q_1=\frac{1}{\sqrt{2}}\left(-i\frac{d}{dx}\sigma_1
-W(x)\sigma_2\right),\quad
Q_2=i\sigma_3Q_1,
\label{qq}
\end{equation}
\begin{equation}
H=\frac{1}{2}
\left(-\frac{d^2}{dx^2}+W^2(x)-W'(x)\sigma_3\right).
\label{ham}
\end{equation} 
Let us try to do this
in a way that respects
the rotational $J_3^2$-symmetry of the spin space. 
The list of nontrivial independent operators 
respecting $J_3^2$ symmetry 
is given by 
the operators commuting or
anticommuting with $J_3$,
\[
{\Sigma}_1=\frac{1}{2}(\sigma_3\otimes1-1\otimes\sigma_3),\quad
\Sigma_2=\frac{1}{2}
(\sigma_1\otimes\sigma_2-
\sigma_2\otimes\sigma_1),\quad
\Sigma_3=\frac{1}{2}(\sigma_1\otimes\sigma_1+
\sigma_2\otimes\sigma_2),
\]
\[
\Xi_1=\frac{1}{2}(\sigma_1\otimes\sigma_1-
\sigma_2\otimes\sigma_2),\quad
\Xi_2=\frac{1}{2}
(\sigma_1\otimes\sigma_2+
\sigma_2\otimes\sigma_1),
\]
and by $J_3\equiv\Xi_3$ itself.
Operators $\Sigma_i$ annihilate the states $\chi^+$, $\chi^-$,
i.e. they are proportional effectively to  $1-J_3^2$
which projects on the $j_3=0$ subspace, whereas 
operators $\Xi_i$,
being proportional to $J_3^2$,
annihilate the states $\chi_s$, $\chi_a$.
Rearranging $4$-dimensional spin space into the 
direct sum of $j_3=0$ and $j_3^2=1$ subspaces,
we find that in the basis 
$(\chi_s,\chi_a){\oplus}
(\chi^+,\chi^-)$
the action of the listed operators 
can be represented as
\begin{equation}
\Sigma_i=\sigma_i{\oplus}(1-J_3^2),\quad
\Xi_i=J_3^2{\oplus}\sigma_i,
\label{sxi}
\end{equation}
i.e. $\Sigma_1\chi_{s(a)}=\chi_{a(s)}$, 
$\Sigma_3\chi_{s(a)}=+(-)\chi_{s(a)}$,
$\Sigma_i\chi^\pm=0$,
$\Sigma_2=i\Sigma_1\Sigma_3$,
$\Xi_1\chi^\pm=\chi^\mp$, $\Xi_3\chi^\pm=\pm\chi^\pm$,
$\Xi_i\chi_{s(a)}=0$,
$\Xi_2=i\Xi_1\Xi_3$.

The `physical' operators are those 
transforming the states of the form
(\ref{gen}) into the states of the same form
(i.e. one should remember that we are
dealing with the system of two identical fermions).
The algebra of physical operators
is generated by 
\begin{equation}
{\cal A}_+=f_+({x},\frac{d}{d{x}}){\cal O}_+,\quad
{\cal A}_-=f_-({x},\frac{d}{d{x}}){\cal O}_-,
\label{phys}
\end{equation}
where 
$f_\pm(-{x},-\frac{d}{d{x}})=
\pm f_\pm({x},\frac{d}{d{x}})$,
${\cal O}_+=1,\Sigma_3,\Xi_i$, $i=1,2,3$,
and ${\cal O}_-=\Sigma_1,
\Sigma_2$.

Taking into account the explicit form of the operators
(\ref{sxi}),
we see that the  supercharge operators cannot take
the form (\ref{qq}) on the subspace with
$j_3^2=1$ since the differential operator
$\frac{d}{dx}$ is odd.
On the other hand, defining 
the operators
\begin{equation}
{\cal Q}_1=\frac{1}{\sqrt{2}}\left(-i\frac{d}{dx}{\Sigma_1}
-W_-(x)\Sigma_2\right),\quad
{\cal Q}_2=i\Sigma_3{\cal Q}_1,
\label{cqq}
\end{equation}
\begin{equation}
{\cal H}=\frac{1}{2}
\left(\left(-\frac{d^2}{dx^2}+W^2_-(x)\right)
{\cal I}-W'_-(x)\Sigma_3\right)
\label{cham}
\end{equation}
where $W_-(x)$ is an odd function, and
${\cal I}=1\oplus (1-J_3^2)$,
we find that they satisfy $N=1$ superalgebra,
$\{{\cal Q}_a,{\cal Q}_b\}=2\delta_{ab}{\cal H},$
$[{\cal H},{\cal Q}_a]=0$. On the subspace $j_3=0$,
these operators take 
the form of operators (\ref{qq}),
(\ref{ham}).
Moreover, since these operators annihilate $j_3^2=1$ states,
we conclude that the Hamiltonian (\ref{cham})
and the supercharges (\ref{cqq}) 
provide an $N=1$ supersymmetry
realized on the system of two identical
fermions.
Spin states with $j_3=1$ and $j_3=-1$ 
are supersymmetric vacuum states with zero energy
regardless of whether supersymmetry
in $j_3=0$ sector is exact or spontaneously broken.

The only essential difference of the supersymmetry 
realized in the sector $j_3=0$ 
from Witten's SUSYQM
is that in the two-fermion system
the superpotential $W_-(x)$ must be 
an odd function. But we have 
exactly the same restriction
in the case of minimally
bosonized supersymmetric quantum mechanics 
and one can establish the one-to-one correspondence
of supersymmetric $j_3=0$ subsystem
with the bosonized supersymmetric system.
To this end, 
let us consider another system described by
the scalar wave function 
${\Psi}(x)={\Psi}_+(x)+{\Psi}_-(x)$,
${\Psi}_\pm(-x)=\pm {\Psi}_\pm(x)$.
The subspaces formed by even and odd functions 
are mutually orthogonal,
they can be distinguished by 
the parity operator $R$, $R {\Psi}(x)={\Psi}(-x)$:
$R{\Psi}_\pm(x)=\pm {\Psi}_\pm(x)$.
Even operators $f_+(x,\frac{d}{dx})$ map these two subspaces into
themselves,
whereas odd operators $f_-(x,\frac{d}{dx})$
interchange these two subspaces.
Taking into account that $\Sigma_2=i\Sigma_1\Sigma_3$, we arrive at the
one-to-one correspondence between
the $N=1$ supersymmetry realized on the system of two identical fermions
in the $j_3=0$ sector
and the bosonized supersymmetry 
through the following identifications:

\vspace{0.5cm}

\begin{tabular}{|l|c|c|} 
\hline
        & Fermion system ($j_3=0$)  &  Bosonized SUSYQM \\ 
  \hline
States    & $\chi_s\psi_-$           &   ${\Psi}_-$  \\
          & $\chi_a\psi_+$           &   ${\Psi}_+$  \\ \hline
Operators & $f_+{\cal I}$            &   $f_+$       \\ 
          & $f_+{\Sigma_3}$          &   $f_+R$      \\
          & $f_-{\Sigma_1}$          &   $f_-$       \\ 
    & $f_-{\Sigma_2}$          &   $if_-R$     \\ \hline
\end{tabular} 
\vspace{0.5cm}

The $N=1$ supersymmetry can also be realized in the  
two-fermion system in 2 dimensions.
In this case the supercharges and the Hamiltonian 
are 
\begin{equation}
{\cal Q}_1=\frac{1}{\sqrt{2}}\pi^a_-\Sigma_a,
\quad
{\cal Q}_2=i\Sigma_3{\cal Q}_1,\quad
{\cal H}=\frac{1}{2}(\pi_-^{a2}{\cal I}
-\Sigma_3 B_+),
\label{dim2}
\end{equation}
where $\pi^a_-=-i\partial^a-A^a_-({\bf x})$,
$\partial^a=\partial_ a=\partial/\partial x^a$,
$a=1,2$, ${\bf A}_-({\bf x})$
is a two-dimensional antisymmetric vector potential, 
${\bf A}_-(-{\bf x})=-{\bf A}_-({\bf x})$,
and the magnetic field
$B_+({\bf x})=\partial_1 A^2_-({\bf x})-
\partial_2A_-^1({\bf x})$ is symmetric under inversion
${\bf x}\rightarrow -{\bf x}$.
In this case the states
$j_3=\pm 1$ 
are also
supersymmetric vacuum states with zero energy. 
The $N=1$ supersymmetry realized in the $j_3=0$ subspace
of two-fermion system
corresponds formally to SUSYQM
of the two-dimensional (plane) 
spin-$1/2$ particle
with gyromagnetic ratio $g=2$
interacting with a magnetic field
given by a generic vector potential ${\bf A}({\bf x})$ 
\cite{d2,susyqm}
(there is no parity restriction on the vector gauge potential): 
\begin{equation}
Q_1=\frac{1}{\sqrt{2}}\pi^a\sigma_a,\quad
Q_2=i\sigma_3Q_1,\quad
H=\frac{1}{2}(\pi^{a2}-\sigma_3 B).
\label{d2}
\end{equation}
The $N=1$ supersymmetry of the
$j_3=0$ subspace of the system (\ref{dim2}) 
gives rise to the bosonized 2-dimensional
supersymmetric system in the same way
as we indicated in the one dimensional case
with the only difference that 
the reflection (parity) operator
$R$ is now given by $R\equiv R_1R_2$,
where $R_1$, $R_2$ are
reflection operators
with respect to $x^1$ and $x^2$:
$\{R_1,x^1\}=\{R_2,x^2\}=0$,
$[R_1,x^2]=[R_2,x^1]=0$,
$R^2_1=R^2_2=1$, $[R_1,R_2]=0$. 
The operator $R$ is also
the operator of space rotation by the angle $\pi$,
$R=\exp(-i\pi L)$, where $L=i(x^2\partial_1-x^1\partial_2)$
is the operator of orbital angular momentum.

It is known that the conventional
$N=1$ SUSYQM
can be constructed in 3-dimensional coordinate
space in the form analogous to that of 2-dimensional space
(\ref{d2}), but this time
the vector gauge potential must be
an antisymmetric function, 
${\bf A}({\bf x})=-{\bf A}(-{\bf x})
\equiv {\bf A}_-({\bf x})$.
In this case, the $N=1$ supercharge and Hamiltonian operators 
have the form \cite{d2}
\begin{equation}
Q_1=\frac{1}{\sqrt{2}}\pi_-^j\sigma_j,\quad
Q_2=iPQ_1,\quad
H=\frac{1}{2}(\pi_-^{j2}-B_+^j\sigma_j),
\label{d3}
\end{equation}
where $B^j_+=\epsilon_{jkl}\partial_k A_-^l$
is a supersymmetric pseudovector of magnetic field,
and $P$ is the parity operator,
$\{x^i,P\}=0$, $i=1,2,3$, $P^2=1$.
Since in the supercharge $Q_1$
all $\sigma$-matrix factors are multiplied by
odd operators $\pi_-^j$, it is clear
that this 3-dimensional supersymmetry
cannot be reproduced in the two-fermion system:
the reason is that the form of the physical operators
(\ref{phys}) requires that  $\Sigma_3$ should be 
multiplied only by even space operator.
As a consequence, it is not possible
to construct the bosonized analog of the supersymmetry
(\ref{d3}) in the way described in Section 2.

\section{Reflection-dependent  unitary transformation
applied to Dirac field theory}

It is well known that
Witten's SUSYQM is related in some aspects
to the Dirac field theory \cite{d2,susyqm,dirac}
(see, e.g., Section 4.3 below).
With this motivation,
here we apply the special unitary transformation
(\ref{unop})
to the $2D$ Dirac field theory.
First, we find that the transformation
diagonalises the Hamiltonian operator of the $2D$ free massive 
Dirac field.  Unlike the Foldy-Wouthuysen case, the
resulting Hamiltonian
is not of a square root form but is linear in space derivative
and contains a space reflection (parity) operator.  Subsequent
reduction to `up' or `down' field component breaks the
Poincar\'e invariance of the theory, but supplies a
linear differential equation with reflection 
whose `squared form' is
the massive Klein-Gordon equation.
In the massless limit 
this linear equation becomes
the self-dual Weyl equation.  
Then we show
that the linear differential equation with reflection admits the
generalization to higher dimensions and 
also allows gauge interactions.
Finally,
we show that the 
bosonized $N=1$ SUSYQM
emerges if the nonlocal
unitary transformation is applied
to the Dirac field theory
in a kink background.

\subsection{Linear differential equation with reflection}

Let us consider free massive Dirac equation in $1+1$ dimensions,
\[
[i(\gamma^0\partial_t+
\gamma^1\partial_x) -m]\Psi(t,x)=0,
\]
and apply to it the unitary transformation 
\begin{equation}
\Psi(t,x)\rightarrow \Psi'(t,x)=U\Psi(t,x),\quad
U=\exp\left[{i}\pi{\cal S}_-\Pi_+\right]=
{\cal S}_+-R{\cal S}_-.
\label{operu}
\end{equation}
Here ${\cal S}_\pm=\frac{1}{2}(1\pm\gamma_5),$
$\gamma_5=\gamma^0\gamma^1$,
$\Pi_+=\frac{1}{2}(1+R)$, and
$R$ is the space reflection (parity) operator,
$Rt=tR$, $Rx=-xR$, $R^2=1$.
Transformation (\ref{operu}) 
is generated by the operator which is
the product of the projector 
on the subspace of even functions, 
$\Pi_+\Psi(t,x)=\Psi_+(t,x)$,
$\Psi_+(t,x)=\Psi_+(t,-x)$,
and of the chiral projector 
${\cal S}_-$.
In representation 
$\gamma^0=\sigma_3$, $\gamma^1=i\sigma_2$, 
operator $U$ is reduced to the 
unitary operator introduced in Section 2.
With this unitary transformation,
$A\rightarrow A'=UAU^{-1}$. For
$A=t,x,\sigma_i$, $i=1,2,3$, this means
$t\rightarrow t'=t$, $x\rightarrow x'=
\sigma_1 x$, $\sigma_1\rightarrow \sigma'_1=\sigma_1$,
$\sigma_2\rightarrow \sigma'_2=-R\sigma_2,$
$\sigma_3\rightarrow \sigma'_3=-R\sigma_3$.
Multiplying  the transformed Dirac equation by $\sigma_3$,
we arrive at
\begin{equation}
[iR(\partial_t+\partial_x)+m\sigma_3]\Psi'(t,x)=0.
\label{peq}
\end{equation}
Thus, the transformed Dirac field $\Psi'$ satisfies the
equation
$i\partial_t\Psi'=H'\Psi'$
with the Hamiltonian 
\begin{equation}
H'=-i\partial_x-R\sigma_3 m.
\label{hp}
\end{equation}
The Hamiltonian (\ref{hp})
has diagonal matrix form and 
therefore (\ref{operu})
is analogous to the Foldy-Wouthuysen (FW) transformation.
However, the
Hamiltonian (\ref{hp}) is not of the FW square root form,
but the relation
\[
H'{}^2=-\partial_x^2+m^2
\]
is satisfied here due to the 
dependence of the Hamiltonian
on the reflection operator $R$.
Like the FW case, the transformation 
(\ref{operu}) is nonlocal:
the transformed field at point $x$ is a
linear combination of the positive chirality
component of the field at point $x$ and 
of the negative chirality component taken at point $-x$:
\[
\Psi'(t,x)={\cal S}_+\Psi(t,x)-{\cal S}_-\Psi(t,-x).
\] 

Since the transformed Dirac equation (\ref{peq}) has a diagonal
form, one could reduce the theory to that of either up, 
$\Psi'_1\equiv
\psi^{+}$,
or down, $\Psi'_2\equiv \psi^{-}$, field component, 
each of which 
satisfies the
corresponding linear differential equation,
\begin{equation}
[iR(\partial_t+\partial_x)+\epsilon m]\psi^{\epsilon}=0,
\quad \epsilon=+,-.
\label{PE}
\end{equation}
In the light-cone coordinates
$x_\pm=t\pm x$,  Eq. (\ref{PE}) is
\[
i\partial_+\psi^\epsilon (x_+,x_-)+\epsilon m\psi^\epsilon(x_-,x_+)=0.
\]
Both fields $\psi^+$ and
$\psi^-$ satisfy the massive Klein-Gordon equation
as a consequence of 
equation (\ref{PE}), which we call 
{\it the linear differential equation with reflection}.
However, the reduction of the Dirac field theory
given by Eq. (\ref{peq})
to the one-component field theory corresponding to Eq. (\ref{PE})
destroys the Poincar\'e invariance.
This is clear from the form of the space
translation and Lorentz transformations,
\begin{equation}
\delta_\kappa\Psi'=\kappa\cdot \sigma_1\partial_x\Psi',
\quad
\delta_\lambda\Psi'=\lambda\cdot\sigma_1
\left(t\partial_x+x\partial_t-
\frac{1}{2}\right)\Psi',
\label{tl}
\end{equation}
which mixes the components $\Psi'_1=\psi^+$ and $\Psi'_2=\psi^-$.
Here $\kappa$ and $\lambda$ are the 
corresponding infinitesimal transformation parameters.
It is interesting to note that
in spite of this breaking of Poincar\'e invariance,
the theory given by Eq. (\ref{PE})
has a formal analogy with the initial,
non-transformed Dirac equation.
Indeed, decomposing the field $\psi^\epsilon$ in even and odd parts,
$\psi^\epsilon=\psi^\epsilon_++\psi^\epsilon_-$,
$\psi^\epsilon_\pm(t,-x)=\pm\psi^\epsilon_\pm(t,x)$,
we represent one equation (\ref{PE}) with $\epsilon=+$ or $-$
in the form of two equations:
\[
i(\partial_t\psi^\epsilon_++\partial_x\psi^\epsilon_-)+
\epsilon m\psi^\epsilon_+=0,\quad
-i(\partial_t\psi^\epsilon_-+\partial_x
\psi^\epsilon_+)+\epsilon m\psi^\epsilon_-=0.
\]
These two equations resemble
the Dirac equation 
$[i(\gamma^0\partial_t+\gamma^1\partial_x)+\epsilon m]\Psi^\epsilon=0$
written in terms of 
`up', $\Psi^\epsilon_1$, and `down',  $\Psi^\epsilon_{2}$,
components,
i.e. {\it formally} $\psi^\epsilon_+$ and $\psi^\epsilon_-$
play  the role analogous to
$\Psi^\epsilon_1$ and $\Psi^\epsilon_2$,
respectively. 

The broken Poincar\'e invariance can be `restored' 
in the $m=0$ limit. 
To do this, first
one can note that in this limit Eq. (\ref{PE}) is
the massless self-dual equation for a one-component
Weyl fermion: $(\partial_t+\partial_x)\psi=0$ 
\cite{fj}. For this field, the space-time translation
and Lorentz transformations have the infinitesimal form
\begin{equation}
\delta_\tau\psi=\tau\partial_t\psi,\quad
\delta_\kappa\psi=\kappa\partial_x\psi,\quad
\delta_\lambda\psi=\lambda
\left(t\partial_x +  x\partial_t-\frac{1}{2}\right)\psi,
\label{poimod}
\end{equation}
and the corresponding theory is Poincar\'e invariant.
So, if one ignores the connection of the theory
given by Eq. (\ref{PE}) with the initial massive Dirac
theory and
postulates for the field $\psi^\epsilon$ 
the same form of Poincar\'e transformations (\ref{poimod}), 
then in the corresponding field action
${\cal S}=\int{\cal L}d^2x$ with
\begin{equation}
{\cal L}^\epsilon=\bar{\psi^\epsilon}[iR(\partial_t+\partial_x)+\epsilon
m]\psi^\epsilon,\quad \bar{\psi}=\psi^\dagger R,
\label{l1+1}
\end{equation}
it is the mass term that breaks space translation and Lorentz
invariance (but does not break time translation invariance).
In this way, the field 
satisfying the linear differential equation with reflection
(\ref{PE})
can be viewed as a massive Poincar\'e non-invariant
generalization of the Poincar\'e invariant theory
of massless Weyl field.
 
\subsection{Higher-dimensional generalization
and switching on interactions}

The Lagrangian (\ref{l1+1}) can be generalized to the case of
$(2+1)$ dimensions:
\begin{equation}
{\cal L}^\epsilon =\bar{\psi}{}^\epsilon L^\epsilon \psi^\epsilon,\quad
L^\epsilon=\Delta +\epsilon m,\quad
\Delta=iR_2[R_1(\partial_0+\partial_1)+\partial_2],
\label{ld2}
\end{equation}
where 
$\bar{\psi}{}^\epsilon=\psi^{\epsilon\dagger} R_1R_2,$
$\epsilon=+,-$, 
$R_1^2=R^2_2=1$,
$R_1R_2=R_2R_1$,
$R_1x_1=-x_1R_1$,
$R_2x_2=-x_2R_2$,
$R_1x_2=x_2R_1$, $R_2x_1=x_1R_2$,
$R_ix_0=x_0R_i$, $i=1,2$.
The equation of motion $L^\epsilon\psi^\epsilon(x_0,x_i)=0$
is a linear differential equation whose 
`square' is the $D=2+1$ 
Klein-Gordon equation:
multiplying $L^\epsilon$
by the operator
$\Delta-\epsilon m$, and using the relation
$\Delta^2=-\partial_0^2+\partial_1^2+\partial_2^2$,
we get the latter equation.
The  generalization to the arbitrary case of $D=d+1$
is achieved via the following formal substitutions
in the corresponding 
$(d+1)$-dimensional Dirac equation:
\begin{eqnarray}
&\gamma^d\partial_d\rightarrow 
R_d \partial_d,\quad
\gamma^{d-1}\partial_{d-1}\rightarrow
R_d R_{d-1}\partial_{d-1},\ldots,\quad
\gamma^{2}\partial_2\rightarrow
R_d R_{d-1}\ldots R_2\partial_2,&
\nonumber\\
&\gamma^{1}\partial_1\rightarrow
R_d R_{d-1}\ldots R_1\partial_1,\quad
\gamma^{0}\partial_0\rightarrow
R_d R_{d-1}\ldots R_1\partial_0.&
\label{gamr}
\end{eqnarray}
The construction admits an arbitrary permutation of spatial indices.
 
Let us return to the transformed Dirac equation 
(\ref{peq}) in order to generalize the linear differential
equation with reflection
for the case of interaction.
We start from the $(1+1)$-dimensional 
Dirac equation minimally coupled to the U(1)
gauge field:
$[i\gamma^\mu D_\mu-m]\Psi=0$,
$D_\mu=\partial_\mu-ieA_\mu$.
Applying to it the same transformation as in a free case,
we get
\begin{equation}
[iR({\cal D}_0+{\cal D}_1)+\sigma_3 m+
ie\sigma_2(A_{0-}+A_{1+})]\Psi'=0,
\label{int1}
\end{equation}
where ${\cal D}_0=\partial_t-ieA_{0+}$, 
${\cal D}_1=\partial_x-ieA_{1-}$,
and $A_\mu=A_{\mu+}+A_{\mu-}$,
$A_{\mu\pm}(t,-x)=\pm A_{\mu\pm}(t,x)$.
In the case when the electric field
$F=\partial_0 A_1-\partial_1 A_0$
is such that its even part is zero,
$F_+(t,x)=~0$,
the non-diagonal term can be removed from Eq. (\ref{int1})
by the gauge transformation 
$A_\mu\rightarrow A_\mu+\partial_\mu \Lambda$
with $\Lambda(t,x)=-\int^x_0 A_{1+}(t,x')dx'+\lambda(t)$.
As a result the equation takes the diagonal form
\begin{equation}
[iR({\cal D}_0+{\cal D}_1)+\sigma_3 m]\Psi=0.
\label{int2}
\end{equation}
The quadratic equation following from
Eq. (\ref{int2}) is
\begin{equation}
[{\cal D}_\mu{\cal D}^\mu+ie {\cal F}-m^2]\Psi'=0,
\label{kg1}
\end{equation}
where ${\cal F}=F_-$. 
In this case the transformed gauge potential components
satisfy the following parity restriction
conditions: $A_0(t,-x)=A_0(t,x)$,
$A_1(t,-x)=-A_1(t,x)$,
i.e. $A_0$ and $A_1$
obey the same (anti)commutation relations
with space reflection operator $R$ as
the time, $t$, and space, $x$.
Reducing Eq. (\ref{int2}), 
we find the generalization of Eq. (\ref{PE})
for the case of $U(1)$ interaction:
\begin{equation}
[iR({\cal D}_0+{\cal D}_1)+\epsilon m]\psi^\epsilon=0,\quad
\epsilon=+,-,
\label{pea}
\end{equation}
where $\psi^+=\Psi'_1$, $\psi^-=\Psi'_2$,
${\cal D}_0=\partial_t-ieA_{0}$, 
${\cal D}_1=\partial_x-ieA_{1}$,
with $A_0$ and $A_1$ being even and odd functions
of $x$, respectively.
Eq. (\ref{pea}) leads to the second order equation
for the field $\psi^\epsilon$
of the form (\ref{kg1}).
Note that as in a free case,
equation (\ref{pea})
and the corresponding second order equation 
decomposed in terms of even and odd parts of $\psi^\epsilon$,
$\psi^\epsilon=\psi^\epsilon_+
+\psi^\epsilon_-$, are
formally analogous to Dirac equation
$
(i\gamma^\mu D_\mu+\epsilon m)\Psi^\epsilon=0
$
and associated Klein-Gordon equation
$(D_\mu D^\mu+ieF-m^2)\Psi=0$, 
decomposed in `up' and `down' components.

Introducing the corresponding restrictions on the 
gauge potential, one can generalize
 the free equation (\ref{PE})
to higher-dimensions with a U(1) interaction. 
For $D=2+1$, this is achieved by imposing the following parity
restrictions:
$A_0(x_0,x_1,x_2)=
A_0(x_0,-x_1,x_2)=A_0(x_0,x_1,-x_2)$,
$A_1(x_0,x_1,x_2)=-A_1(x_0,-x_1,x_2)=
A_1(x_0,x_1,-x_2)$,
$A_2(x_0,x_1,x_2)=A_2(x_0,-x_1,x_2)=
-A_2(x_0,x_1,-x_2)$.
In this case the associated quadratic 
equation is given by 
\begin{equation}
[{\cal D}_\mu{\cal D}^\mu
+ie({\cal F}_{01}+({\cal F}_{02}-{\cal F}_{12})R_1)-m^2]\psi^\epsilon=0,
\label{kgd2}
\end{equation}
where ${\cal F}_{\mu\nu}=
\partial_\mu A_\nu-\partial_\nu A_\mu$.
In order to include the U(1) gauge interaction in arbitrary
$D=d+1$, we have to introduce the obvious parity restrictions for the
components of the vector gauge potential $A_\mu$
generalizing the $D=2+1$ case: 
the components have to satisfy the same
(anti)commutation relations with the set of $d$ reflection
operators $R_i$, $i=1,\ldots,d$, as the coordinates $x_\mu$ do.

The described construction also works for non-Abelian
gauge interaction. It is sufficient to take the field 
$\psi^\epsilon(t,{\bf x})$ in the
corresponding representation of the internal gauge group, and
impose on the algebra-valued vector gauge potential the same
parity restrictions, as in the case of the U(1) interaction.

\subsection{Dirac field in a kink background and
bosonized SUSYQM} 

Conventional SUSYQM underlies the dynamics of Dirac field
propagating in a background of a stationary
scalar field soliton \cite{susyqm}.  
Here we show that the bosonized SUSYQM can also
be revealed for a Dirac field in a kink background.
Analogously to the case of Witten's supersymmetry, the exact or
broken phases of minimally bosonized SUSYQM may be associated
with the presence or absence of fermionic zero modes in a system.

To see this, let us consider the Lagrangian 
\begin{equation}
{\cal L}={\cal L}_s
+\bar{\Psi}[i\gamma^\mu\partial_\mu+gW(\phi)]\Psi,
\label{scalr}
\end{equation}
with ${\cal L}_{s}=
\frac{1}{2}(\partial_\mu\phi)^2-V(\phi)$
describing nonlinear scalar field $\phi$ interacting 
with Dirac field $\Psi$.
The Lagrangian (\ref{scalr}) 
was used in the context 
of fermion number fractionalization \cite{jr}
observed in certain polymers like
polyacetylene, and also 
in supersymmetric field
theories in $1+1$ \cite{bf} (for $V=\frac{1}{2}v^2$ 
and $gW=v'$ with $v=v(\phi)$ being a superpotential).

Applying the unitary transformation (\ref{operu})
to the Dirac field $\Psi$,
the fermion part becomes 
\[
{\cal L}_f=\bar{\Psi}{}'[iR(\partial_t+\partial_x)
-g\sigma_3W_+(\phi)-ig\sigma_2W_-(\phi)]\Psi',\quad
\bar{\Psi}{}'=\Psi'{}^\dagger R,
\]
where we have used the decomposition 
$W(\phi)=W_+(\phi)+W_-(\phi)$,
$W_\pm(\phi(t,x))=\pm W_\pm(\phi(t,-x))$.
With an additional 
unitary transformation $\Psi'\rightarrow \psi=U'\Psi'$
generated by 
$U'=\exp(i\frac{\pi}{4}\sigma_1)$,
we get finally for the Lagrangian (\ref{scalr}) 
\begin{equation}
{\cal L}\rightarrow
{\cal L}={\cal L}_s+
\bar{\psi}[iR(\partial_t+\partial_x)+
g\sigma_2 W_+(\phi)-ig\sigma_3 
W_-(\phi)]\psi,
\quad
\bar{\psi}=\psi^\dagger R.
\label{dbos}
\end{equation}  
Suppose now that in the free case
($g=0$), the nonlinear scalar field theory has a
static kink solution $\phi_c$, satisfying the equation
$d^2\phi_c(x)/dx^2=dV(\phi)/d\phi
\vert_{\phi=\phi_c}$,
and require that $W(\phi)$ is such that
$W_+(\phi_c)=0$.
To see that this requirement for $W$ is not
too tough, consider, as an example, 
$\phi^4$ model 
with $V(\phi)=\lambda^2(\phi^2-a^2)^2$, $\lambda>0$,
$a>0$,
and sine-Gordon model with $V(\phi)=\cos\phi$.
These models have kink solutions of the form 
\[
\phi^4:\,
\phi_c(x)=a\tanh\left(\frac{a\lambda}{\sqrt{2}} (x-X_1)\right),\quad
sG:\,
\phi_c(x)=4\tan^{-1}\exp (x-X_2),
\]
where $X_1$, $X_2$ are some constants 
(see, e.g., \cite{fi4}).
For $\phi$ close to corresponding kink solution,
the choice $W(\phi)={\cal W}_-(\Phi(\phi))$,
with arbitrary odd function ${\cal W}_-$
and
\[
\phi^4:\, \Phi(\phi)=
\frac{\sqrt{2}}{a\lambda}\tanh^{-1}(a^{-1}\phi)+X_1,\quad
sG:\, \Phi(\phi)=\ln\tan(\phi/4)+X_2
\]
satisfies the imposed requirement 
for $\phi^4$ and sine-Gordon models.
In the case of $\phi^4$ model,
kink solution with $X_1=0$ admits, obviously,
a simple choice  $W(\phi)=\phi$.

Then, for field $\psi$ propagating in this
static background the equations of motion
take the diagonal form for up, $\psi^+$,
and down, $\psi^-$, components:
\[
[iR(\partial_t+\partial_x)-\epsilon igW_-(\phi_{c})]\psi^\epsilon=0,
\quad
\epsilon=+,-.
\]
Since $\phi_{c}$ is static,
one can factorize $\psi^\epsilon(t,x)$ as
$\psi^\epsilon(t,x)=\exp(-i\omega t)\psi^\epsilon(x)$, 
$\omega=const$.  As a result, 
we arrive at the equation $\hat{H}{}^\epsilon
\psi^\epsilon(x)=
\omega\psi^\epsilon(x)$ with
$\hat{H}{}^\epsilon=-i(\partial_x +\epsilon g W_-(\phi_{c}(x))R)$. 
The Hamiltonian $\hat{H}{}^\epsilon$
has the form of the supercharge of the bosonized 
supersymmetric quantum mechanical system
with the superpotential
$W^\epsilon_-(x)=\epsilon gW_-(\phi_{c}(x))$:
\[
\hat{H}{}^\epsilon=Q^\epsilon_1=
-i\left(\frac{d}{dx}+W^\epsilon_-(x)R\right).
\]
The bosonized supercharges $Q_1^\epsilon$ and
$Q^\epsilon_2=iRQ^\epsilon_1$ 
and the operator 
$H^\epsilon=[-\frac{d^2}{dx^2}+W^\epsilon_-{}^2-
\frac{d}{dx}W^\epsilon_- R]$,
form an $N=1$ SUSY algebra:
$\{Q^\epsilon_i,Q^\epsilon_j\}=2\delta_{ij}H^\epsilon$,
$[H^\epsilon,Q^\epsilon_1]=[H^\epsilon,Q^\epsilon_2]=0$.
Only one of the two equations
$\frac{d}{dx}\psi^{\epsilon}(x)=-W^\epsilon_-(x)
\psi^{\epsilon}(x)$, $\epsilon=+,-$,
can have a normalized solution, and if so, $\psi(x)$ is an
even function.  In this case, the corresponding associated
bosonized quantum mechanical system is in the phase of exact
SUSY, and there is a fermionic zero mode solution (corresponding
to $\omega=0$) in the theory given by Lagrangian (\ref{dbos}).
On the other hand, if the equations
$Q^\epsilon_1\psi^{\epsilon}=0$, $\epsilon=+,-$, have no
normalizable solutions, 
the corresponding quantum mechanical
bosonized supersymmetric systems are in a
phase of spontaneously broken
supersymmetry, and the theory (\ref{dbos}) has no fermionic zero
modes.

\section{Discussion and outlook}

To conclude, let us indicate some problems 
that deserve further attention.

Our construction results in even and odd supersymmetry 
generators (\ref{hq12}) given in terms of only bosonic 
operators, i.e. supersymmetry algebra is preserved here.
This is in contrast to the hidden supersymmetry
observed by Gozzi \cite{gozzi}
at the level of the generating functional
of Witten's SUSYQM, where the information
on the supersymmetry  algebra disappears after integrating away 
the anticommuting variables.
On the other hand, at the moment
for us it is not clear how supersymmetry
transformations should be understood in the
bosonized SUSYQM. This is not clear either 
in the case of conventional SUSYQM if the
fermionic operators are realized in matrix form,
without turning to the holomorphic (Grassmann) representation.
Some light on this problem could be shed 
by the observed possibility of interpreting  the minimally bosonized
SUSYQM as supersymmetry of identical fermions.

The minimally bosonized SUSYQM has a
nonlocal nature brought about by the reflection
operator in the unitary transformation 
(\ref{unop}) and in the bosonized supersymmetry generators
(\ref{hq12}).  In this sense, the
construction is analogous to the bosonization of fermionic
theories in $1+1$ dimensions \cite{b1+1} which, in turn, is a
generalization of the one-dimensional Jordan-Wigner transformation   
for spin systems \cite{iw} and its
higher-dimensional extensions \cite{fhz}.  An
open problem we will address elsewhere is constructing the
$(1+1)$-dimensional field analog of the minimally bosonized
SUSYQM which could be realized
in terms of only one scalar field in the simplest case.

According to the constructions realized in Section 2,
two different bosonized supersymmetric systems
given by the superpotentials $W_-(x)$
and $-W_(x)$ together form a system,
unitary equivalent to one 
conventional SUSYQM system with
the superpotential $W_-(x)$ (or~$-W_-(x))$.
On the other hand, it is an open question
what is the conventional supersymmetric system
being equivalent to one bosonized system with superpotential
$W_-(x)$. In the simplest case,
the bosonized supersymmetric system with $W_-(x)=x$ 
is equivalent to the superoscillator
with $W(x)=\pm \sqrt{2}x$, where the 
coefficient $\sqrt{2}$ accounts for the difference
in the energy spectra $E_n$ and $E^+_n$
given by Eqs. (\ref{spe0}) and  (\ref{spe1}).
However, it is not obvious which is
the conventional supersymmetric
system in the phase of spontaneously
broken supersymmetry
possessing the same spectrum $E^-_n$
in (\ref{spe1}),
as the bosonized supersymmetric system 
with $W_-(x)=-x$.

We saw that the superpotential $W_-(x)=x-\frac{\nu}{2x}$
gives the isospectral family of the bosonized supersymmetric
systems specified by the parameter
$\nu>-1$ and containing the bosonized superoscillator system
in the phase of exact supersymmetry ($\nu=0$).
It could be expected that
 bosonized SUSYQM should provide a
universal recipe
for constructing isospectral supersymmetric families.
That is, given a bosonized supersymmetric
system with the superpotential $W_-(x)$,
one could produce other
bosonized supersymmetric systems
isospectral to the first one.

The deformed Heisenberg algebra with reflection 
(\ref{aad}) is closely related to parabosons and parafermions
\cite{para,def}, and we saw that it emerges naturally
in bosonized SUSYQM.
We also showed that 
bosonized SUSYQM can be understood as the supersymmetry
realized in a system of two identical fermions.
Therefore, it would be interesting to 
investigate the relationship between 
the parabosonic and parafermionic systems on one hand,
and the supersymmetric two-fermion systems on other.
Besides, the results of Section 3 seem to indicate that
they could be generalized to relate $n$-particle
spinless integrable models involving exchange operators
\cite{poly} with $n$-particle integrable systems
of identical fermions. 

Conventional supersymmetry plays an
important role in the theory of
$(1+1)$-dimensional integrable systems \cite{susyqm,ros}.  It
would be  interesting to investigate the 
applications of bosonized SUSYQM to
the theory of $(1+1)$-dimensional integrable systems on the
half-line \cite{half}.  One notes also  that 
the $(1+1)$-dimensional linear differential
equation with reflection has some formal analogies with the
theory of a massless boson field on the half-line related, in
turn, via bosonization to the massless fermion
field on the half-line \cite{mint}.
Indeed, in both theories spatial translation
and Lorentz invariance are broken,
whereas the theories are invariant under time translations.
Mixed boundary conditions for the boson field
on the half line contain a parameter of dimension of mass
and they are introduced adding to the 
action functional the mass boundary term.
It is such a term, like mass term in 
our Lagrangian (\ref{l1+1}), that breaks
Lorentz and translation invariance.
An idea analogous to the introduction of a mass parameter
to produce mixed boundary conditions
was used earlier in ref.
\cite{gz} to transform
$(1+1)$-dimensional massive scalar field theory
into relativistic anyon field theory.
Thus, a natural and attractive 
problem would be to quantize the theory
given by Lagrangian (\ref{l1+1}) 
and investigate its relation
to the field theories on the half-line,
especially to the relativistic model of massive
anyons \cite{gz}.
It would be also
interesting to answer the question of
whether the higher dimensional generalizations
of the $(1+1)$-dimensional
linear differential equation with reflection 
can be obtained from the corresponding 
higher dimensional Dirac equation through a
unitary transformation with reduction
analogous to our construction of Section 4.

\vskip1cm
{\bf Acknowledgements}
\vskip5mm

The work was supported in part by 
grants 1960229, 1970151, 1980619 and 1980788 from FONDECYT (Chile)
and by DICYT (USACH).
Institutional support to CECS
from Fuerza A\'erea de Chile and a group of private companies 
(Business Design Associates, CGE,
CODELCO, COPEC, 
Empresas CMPC, Minera Collahasi, Minera Escondida,
NOVAGAS and XEROX-Chile) is also acknowledged.
J.Z. acknowledges support from the J.S. Guggenheim Memorial
Foundation.

\end{document}